\documentclass[conference,a4paper]{IEEEtran}

\usepackage{cite}
\usepackage[pdftex]{graphicx}
\graphicspath{{./images/}}
\usepackage[cmex10]{amsmath}
\usepackage{array}
\usepackage{mdwmath}
\usepackage{mdwtab}
\usepackage{eqparbox}
\usepackage[tight,footnotesize]{subfigure}
\usepackage{caption}
\usepackage[font=footnotesize]{subfig}
\usepackage{stfloats}
\usepackage{url}
\usepackage{epstopdf}

\usepackage{amssymb}

\begin{document}
\title{Improved Channel Estimation for Interference Cancellation in Random Access Methods for Satellite Communications}

\author{\IEEEauthorblockN{Karine Zidane\IEEEauthorrefmark{1},
J\'er\^ome Lacan\IEEEauthorrefmark{1},
Marie-Laure Boucheret\IEEEauthorrefmark{3}, 
Charly Poulliat\IEEEauthorrefmark{3},}
\IEEEauthorblockA{\IEEEauthorrefmark{1}Univesity of Toulouse, ISAE/DMIA \& T\'eSA\\ Email: \{karine.zidane, jerome.lacan\}@isae.fr}
\IEEEauthorblockA{\IEEEauthorrefmark{3}University of Toulouse, IRIT/ENSEEIHT\\ Email: \{marie-laure.boucheret, charly.poulliat\}@enseeiht.fr}
}
\maketitle
\begin{abstract}
In the context of satellite communications, random access methods can significantly increase throughput and reduce latency over the network. The recent random access methods are based on multi-user multiple access transmission at the same time and frequency followed by iterative interference cancellation and decoding at the receiver. Generally, it is assumed that perfect knowledge of the interference is available at the receiver. In practice, the interference term has to be accurately estimated to avoid performance degradation. Several estimation techniques have been proposed lately in the case of superimposed signals. In this paper, we present an overview on existing channel estimation methods and we propose an improved channel estimation technique that combines estimation using an autocorrelation based method and the Expectation-Maximization algorithm, and uses pilot symbol assisted modulation to further improve the performance and achieve optimal interference cancellation.
\end{abstract}
\begin{keywords}
Satellite communication, Network coding, Channel estimation, Expectation-maximization algorithms
\end{keywords}
\section{Introduction}
\label{sec:intro}
Recently, the use of random access (RA) methods in satellite communications has been the center of attention of many researchers. In traditionnal RA methods like Aloha \cite{aloha} and Slotted-Aloha \cite{Saloha}, multiple users transmit their packets simultaneously, but the problem is that the receiver only decodes the contents of clean packets (i.e., packets that did not experience collision) and thus, superimposed packets are ignored by the receiver and retransmitted by the corresponding users. To avoid retransmission delays resulting from the large satellite propagation time, new RA methods like CRDSA (Contention Resolution Diversity Slotted Aloha \cite{CRDSA}) and MuSCA (Multi-Slot Coded Aloha \cite{MUSCA}), have emerged as a solution to this problem. These methods enable the receiver to decode a certain number of packets and remove them from the frame (interference cancellation) so that other packets become collision-free. 

For instance, in CRDSA, each user sends two or three replicas of his packet on the frame, separated by random delays. At the receiver node, several packets corresponding to different users can arrive at the beginning of the same time slot. The receiver detects the time slots containing clean packets and decodes them successfully. Then, the signals corresponding to the decoded packets and their replicas are reconstructed and removed from the frame. 

On  the other hand, MuSCA performs in a similar way but instead of sending replicas of the same packet, each user first encodes the content of one packet with a strong forward error correcting (FEC) code  then sends several parts of a single code word on the frame. At the receiver node, the decoder combines all the parts of a code word and implies them in the decoding process. Thus, with MuSCA, the decoder is able to retrieve information not only from clean packets but also from packets that experienced collision, because the useful information inside each packet is well protected by the FEC code used.

In both RA methods, interference cancellation at the receiver requires perfect knowledge of the channel parameters that have a noticeable effect on the packets to remove. In reality, the receiver does not have knowledge of the channel state information (CSI), therefore channel parameters have to be accurately estimated. Or else, the interference packets are not correctly removed and residual estimation errors are added to the undecoded packets on the frame.

The problem to be addressed in this paper is the accurate channel estimation for RA methods based on interference cancellation. The main issue is to be able to estimate the channel parameters in the case of superimposed signals, in order to achieve performance as good as the perfect knowledge case.

This challenge has been tackled in some previous publications. To estimate several channels simultaneously, Expectation-Maximization (EM) algorithm in \cite{EM2} uses known orthogonal sequences. In \cite{casini}, another approach uses an autocorrelation based method that derives channel amplitudes and frequency offsets from clean packets. In this paper, we propose a channel estimation scheme that exploits the EM algorithm and focuses on the problem of parameter initialization which has not been taken into account in \cite{EM2}. In a second contribution, we use pilot symbol assisted modulation (PSAM) \cite{PSAM} to refine frequency estimation.

The rest of this paper is organized as follows. In Section \ref{sec:sec2}, we present the system model and the main assumptions of our work. We describe prior related work in Section \ref{sec:sec3}. In Section \ref{sec:sec4}, we propose an improved channel estimation scheme and we present experimental results. We conclude and discuss future work in Section \ref{sec:sec5}.

\section{System Model}
\label{sec:sec2}
\begin{figure}[!t]
\begin{minipage}[b]{1.0\linewidth}
  \centering
  \centerline{\includegraphics[scale=0.5]{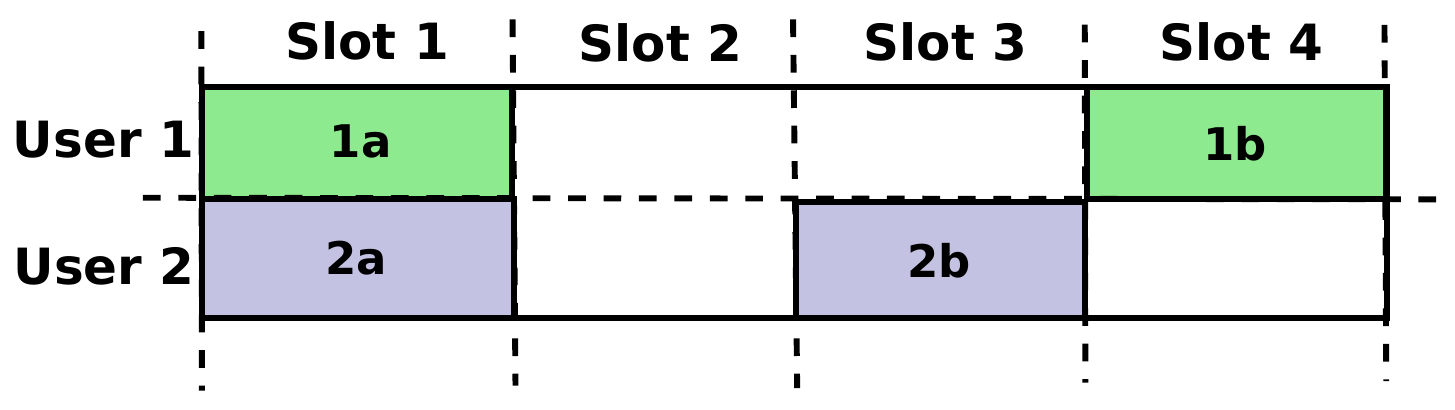}}
\end{minipage}
\caption{Packets transmission with collision on slot $1$}
\label{fig:trame}
\end{figure}

In order to illustrate the main issues raised by imperfect channel estimation in interference cancellation based RA methods, we consider the following example (see  \figurename~\ref{fig:trame}). Each user ($1$ and $2$) sends two replicas ($a$ and $b$) of the same packet on two different time slots on the frame. We suppose that the receiver first detects packet $1b$ as it is a clean packet, decodes it correctly and removes its corresponding signal from Slot~$4$. Then, using the known decoded bits of packet $1b$, the signal corresponding to packet $1a$ is reconstructed and suppressed from Slot~$1$. Thus, packet $2a$ becomes collision free, and has a bigger probability to be decoded successfully.

To correctly remove the signal sent by User~$1$ on Slot~$1$, the receiver needs to estimate the channel parameters associated with this signal. If the channel parameters are not estimated accurately, residual estimation errors are added to the signal of interest sent by User~$2$ on Slot~$1$ and the scheme does not perform well.

\begin{figure}[!t]
\begin{minipage}[b]{1.0\linewidth}
  \centering
  \centerline{\includegraphics[scale=0.5]{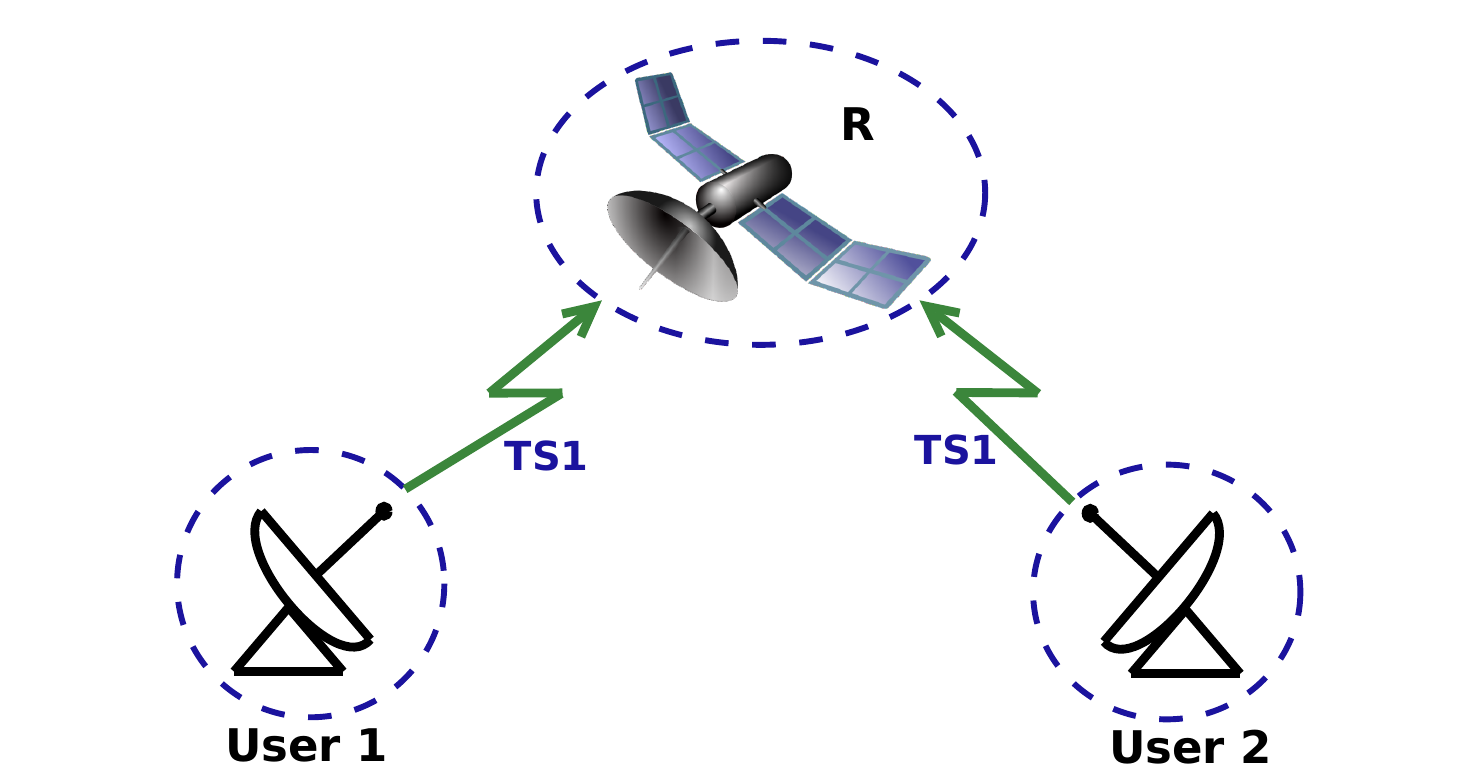}}
\end{minipage}
\caption{Transmitting scenario}
\label{fig:system}
\end{figure}

\begin{figure}[!t]
\begin{minipage}[b]{1.0\linewidth}
  \centering
  \centerline{\includegraphics[scale=0.5]{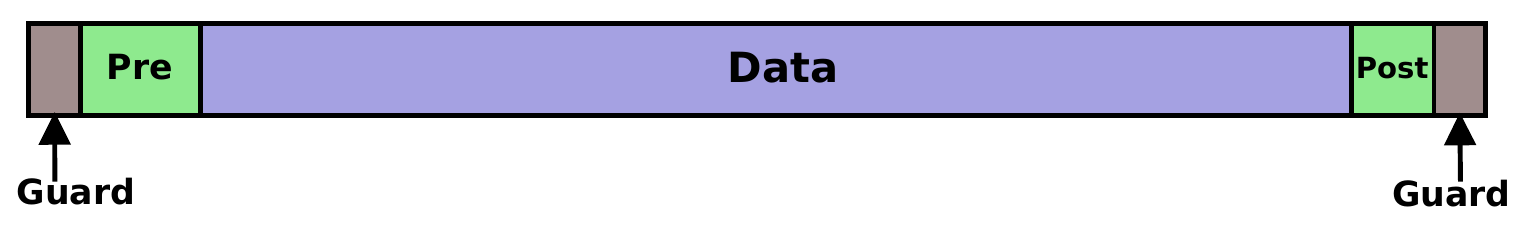}}
\end{minipage}
\caption{Packet strucuture with preamble and postamble}
\label{fig:structure1}
\end{figure}

In the following, we consider a system with two users (User~$1$ and User~ $2$) transmitting their packets to a receiver node $R$ during the same time slot $TS1$, over two different channels $h_1$ and $h_2$ respectively (See \figurename~\ref{fig:system}). We suppose the users synchronized at the symbol level, and we consider an environment where phase noise is neglected. We assume that the receiver knows the number of superimposed packets arriving on the same time slot. The structure of the packet sent by each user is shown in \figurename~\ref{fig:structure1}. Guard intervals are used to delimit the beginning and the end of a packet. The preamble and the postamble are unique orthogonal sequences modulated with binary phase-shift keying (BPSK), known at the receiver node and used for channel estimation.

The received signal, $y(i)$, at the receiver node $R$ during the time slot $TS1$, after pulse shaping, and oversampling by the sampling time $T_e=T_s/Q$, is given by
\begin{equation} \label{eq:received}
y(i)=\sum_{k=1}^{2}h_k(i) \sum _{n=0} ^{L_{p}-1}x_k(n) g(iT_e-nT_s)+w(i)
\end{equation}
where:
\begin{itemize}
\item $i=0,1, ..., QL_{p}-1$ and $n=0,1, ...,L_{p}$ are used to refer to $T_e$-spaced and $T_s$-spaced samples respectively;
\item $x_k(n)$ refers to the $n^{th}$ symbol sent by user $k$;
\item $L_{p}$ is the length in symbols of the entire packet;
\item $T_s$ is the symbol duration;
\item $Q$ is the oversampling factor of the root raised cosine filter;
\item $g$ stands for the root raised cosine pulse function (shaping filter);
\item $w$ is a complex additive white Gaussian noise (AWGN) process;
\end{itemize}
 
We assume a block fading channel model with unknown channel parameters. The channel coefficient $h_k(i)$ having an effect on the signal sent by user $k$ is modeled as given in \cite{EM2}
\begin{equation}
h_k(i)=A_ke^{j(2\pi \Delta f_kiT_e+\varphi _k)}
\end{equation}
where:
\begin{itemize}
\item $A_k$ is a lognormally distributed random variable modeling the channel amplitude of user $k$, assumed to remain constant over the frame duration;
\item $\Delta f_k$ is the frequency offset of the signal sent by user $k$, assumed constant over the frame duration. $\Delta f_k$ is uniformly distributed in $[0,\Delta f_{max}]$ with $\Delta f_{max}$  equal to $1\%$ of the symbol rate $1/T_s$;
\item $\varphi _k$  is the phase shift of the signal sent by user $k$. It is assumed to remain constant over a duration of one time slot. $\varphi _k$ is a random variable drawn independantly from one slot to another from a uniform distribution in $[0, 2\pi]$;
\end{itemize}

Due to prior decoding, we suppose that $R$ already knows the interference symbols $x_1(n)$, sent by User~$1$, and needs to demodulate and decode the signal sent by User~$2$. Therefore, $R$ needs to compute the channel estimates $\widehat{h_1}$ and $\widehat{h_2}$, then suppress the signal corresponding to User~$1$ from $y$ in order to obtain the discrete signal $s_2$ as given below
\begin{align}
s_2(i)&=\widehat{h_2}(i) \sum _{n=0} ^{L_{packet}-1}x_2(n) g(iT_e-nT_s) \nonumber\\  &+(h_1(i)- \widehat{h_1}(i)) \sum _{n=0} ^{L_{packet}-1}x_1(n) g(iT_e-nT_s)\nonumber\\ &+ (h_2(i)- \widehat{h_2}(i)) \sum _{n=0} ^{L_{packet}-1}x_2(n) g(iT_e-nT_s) + w(i) 
\end{align}
In presence of residual channel estimation errors, the signal $s_2$ is matched filtered and sampled with the sampling period $T_s$, and the resulting estimated symbols $s_2(nT_s)$ are given by
\begin{align}
{s_2}(nT_s)&=\widehat{h_2}(nT_s)x_2(n) + \left( h_1(nT_s) - \widehat{h_1}(nT_s)\right) x_1(n) \nonumber\\&+ \left( h_2(nT_s) -\widehat{h_2}(nT_s)\right) x_2(n) + w(nT_s)
\end{align}
Finally, the estimated sequence $s_2$ is demodulated and decoded and its corresponding signal is suppressed from the frame. 

\section{PRIOR RELATED WORK}
\label{sec:sec3}
In this section, we present existing channel estimation methods that might be relevant to solve the problem addressed in this paper.

\subsection{EM estimation}
\label{ssec:sec31}

The EM algorithm \cite{EM} is a two-step iterative estimation method that has been proposed in \cite{EM2} to perform channel estimation in case of superimposed signals. In a first step called ``Expectation'' (E step), the preamble part is extracted from the received signal. Then for each user $k$, vector $p_k$ is derived as the preamble of user $k$ ($pre_k$) multiplied by its estimated channel coefficient $\widehat{h_k}(n)$ (of previous iteration) and additioned with a certain percentage $\beta _k$ of the difference between the received preamble part ($r$) and the reconstructed preambles of both users $1$ and $2$, as shown below
 
\textbf{At the $m^{th}$ iteration,}
\begin{align} \label{eq:EM_eq1}
\widehat{p_k}^{(m)}(n)&= pre_k(n)\widehat{h_k}(n)^{(m-1)}\nonumber\\&+ \beta _k \left[r(n)- \sum _{l=1}^K pre_l(n)\widehat{h_l}(n)^{(m-1)}\right]
\end{align}
\noindent where $n$ refers to the index of the preamble.

In a second step called ``Maximization'' (M step), the mean square error (MSE) between each user's component $p_k$ derived at the E step and a symbol sequence reconstructed using channel parameters to estimate, is minimized as follows

\begin{equation} \label{eq:EM_eq2}
\min _{A',\Delta f ',\varphi '} \sum _{n=0} ^{L_{pre}-1}\left| pre_k(n)\widehat{p_k} ^{(m)}(n)- A'e^{j(2\pi\Delta f 'T_sn+\varphi ')} \right| ^2 
\end{equation}
where $A'$, $\Delta f'$ and $\varphi '$ are tentative values of the channel parameters to be estimated. 

However, with the experimental assumptions considered in our work, the approach used in \cite{EM2} has the following weaknesses:
\begin{itemize}
\item In \cite{EM2}, only preambles at the beginning of each packet are used in the estimation algorithm. However, using grouped training symbols only at the beginning of a packet, makes it difficult to estimate the variable parameters such as the phase of the signal;
\item Channel parameters are initialized randomly in \cite{EM2} at the iteration $m=0$. Nevertheless, random initialization of EM has been proved to be inaccurate in \cite{initial}, and to affect on the correctness of the estimated values;
\end{itemize}

\subsection{Estimation using autocorrelation}
\label{ssec:sec32}

Casini et al. use the method of autocorrelation to estimate channel parameters in \cite{casini}. However, $\Delta f$ is chosen so small that it induces negligible phase variation during a time slot. Moreover, \cite{casini} takes advantage of clean packets in CRDSA to get a good estimation of $A$ and $\Delta f$. Difficulties arise, however, in other random access methods like MuSCA, where finding clean packets in the frame is a rare situation. Note that \cite{casini} presents a system adapted to an environment affected by high phase noise.

\section{Estimation combining EM and autocorrelation}
\label{sec:sec4}

In order to take advantage of the effect of the channel on the transmitted packets at the beginning and the end of a time slot, we find it reasonable to apply the EM algorithm not only on the preamble symbols, as in \cite{EM2}, but also on the postamble symbols. In the case of $K$ colliding packets, EM equations for the $m^{th}$ iteration are:

\subsection*{\textbf{E step}}
\subsubsection*{\textbf{for $k=1,\ldots,K$}}
\begin{align} \label{eq:EM_eq1}
\widehat{p_k}^{(m)}(n)&= b_k(n)\widehat{A_k}^{(m)}e^{j(2\pi\widehat{\Delta f _k}^{(m)}T_sn+\widehat{\varphi _k}^{(m)})}\nonumber\\ &+ \beta _k \left[r(n)- \sum _{l=1}^K b_l(n)\widehat{A_l}^{(m)} e^{j(2\pi\widehat{\Delta f_l}^{(m)}T_sn+\widehat{\varphi _l}^{(m)})}\right]
\end{align}
where:
\begin{itemize}
\item $\widehat{p_k}$ is the estimated preamble concatenated with the estimated postamble of user $k$;
\item $n$ refers to the indexes of the preamble and postamble symbols;
\item $b_k$ is a vector containing BPSK symbols corresponding to the preamble concatenated with the postamble of user $k$;
\item $r$ contains the preamble and the postamble parts of the discrete signal obtained after matched filtering and sampling of the received signal $y$;
\item $\beta _k$ is a coefficient arbitrarily set to $0.8$, for $k=1,\ldots ,K$;
\end{itemize}  

For each iteration $m>0$ of the E step, the values of the channel coefficients $\widehat{A}^{(m)}$, $\widehat{\Delta f}^{(m)}$ and $\widehat{\varphi}^{(m)}$ are equal to the ones obtained at the previous iteration $(m-1)$ of the M step. In \cite{EM2}, at the first iteration $m=0$, the initial values of these parameters are chosen randomly. According to \cite{initial}, random initialization of the channel coefficients in the EM algorithm is not efficient and can lead to undesired results. Therefore, to solve this problem and to speed up the convergence of the EM algorithm, we compute estimates of $\widehat{A}^{(0)}$ and $\widehat{\varphi}^{(0)}$ with the autocorrelation method (Equations \eqref{eq:autocor1} and \eqref{eq:autocor2}) and we use them as initial values at the first iteration of the E step.

\subsection*{\textbf{Initialization by autocorrelation}}
The initial estimated amplitude of the channel $\widehat{A_k}^{(0)}$ corresponding to user $k$ is derived as follows
\begin{equation} \label{eq:autocor1}
\widehat{A_k}^{(0)}= \sum _{n=0} ^{L{pre}+L_{post}-1}\dfrac{r_{pre/post}(n)\times b_k(n)}{L_{pre}+L_{post}}
\end{equation}
and the initial estimated phase offset $\widehat{\varphi _k}^{(0)}$ is calculated as shown below
\begin{equation} \label{eq:autocor2}
\widehat{\varphi _k}^{(0)}= arg(r_{pre/post}\times b_k^T)
\end{equation}
where:
\begin{itemize}
\item $r_{pre/post}$ is a vector containing the received symbols of the preambles and postambles;
\item $b_k^T$ denotes the transpose of vector $b_k$;
\item $\times$ refers to the vector multiplication operator;
\end{itemize}  

\subsection*{\textbf{M step}}
\subsubsection*{\textbf{for $k=1,\ldots,K$}}
\begin{equation} \label{eq:EM_eq2}
\min _{A',\Delta f ',\varphi '} \sum _{n=1} ^{T}\left| b_k(n)\widehat{p_k} ^{(m)}(n)- A'e^{j(2\pi\Delta f 'T_sn+\varphi ')} \right| ^2 
\end{equation}
where:
\begin{itemize}
\item $A'$, $\Delta f'$ and $\varphi '$ are tentative values of the channel parameters to be estimated;
\item T is the vector of preamble and postamble indexes, as $T=\lbrace L_{guard}+1,\ldots ,L_{guard}+L_{pre}, L_{guard}+L_{pre}+L_{pay}+1,\ldots ,L_{guard}+L_{pre}+L_{data}+L_{post}\rbrace$;
\end{itemize}

We evaluate the performance of the previously mentioned estimation techniques, by computing the packet error rate (PER) after demodulating and decoding the discrete signal $s_2$. Without loss of generality, we assume the channel gain of user $k$ $(k=1,...,K)$, $A_k$ normalized to one. We use as preambles and postambles Walsh-Hadamard words of length $80$ and $48$ symbols respectively. The number of iterations needed to achieve convergence of the results of the EM algorithm is three.

\begin{figure}[t]
\begin{minipage}[b]{1.0\linewidth}
  \centering
  \centerline{\includegraphics[scale=0.7]{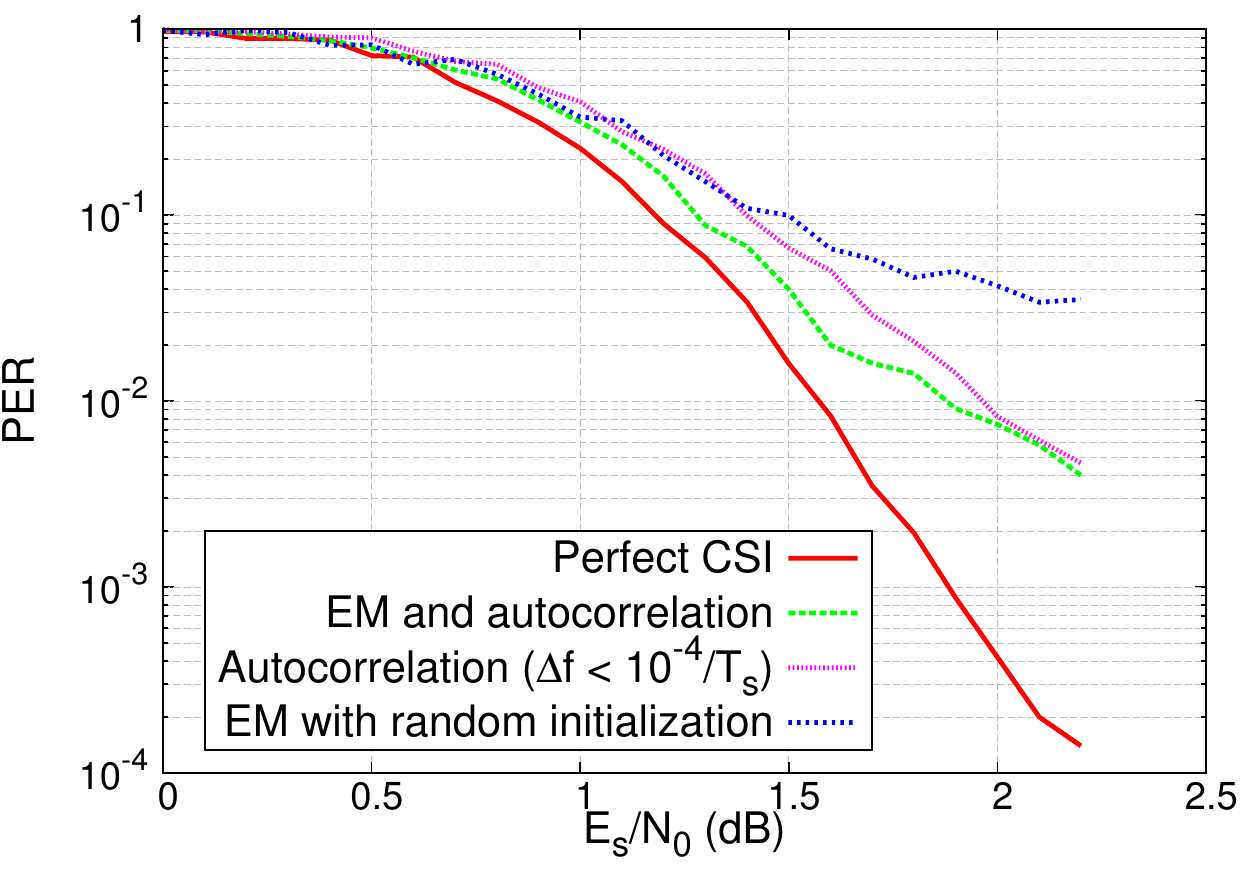}}
\end{minipage}
\caption{PER vs $E_s/N_0$ after interference cancellation and channel estimation using EM, autocorrelation, and EM combined with autocorrelation}
\label{fig:graph1}
\end{figure}

\figurename~\ref{fig:graph1} shows the PER achieved with different estimation methods. The plots are obtained using a turbocode of rate $1/2$ and a codeword length equal to $460$ symbols modulated with quadrature phase-shift keying (QPSK).
Note that only for estimation by autocorrelation, $\Delta f$ is supposed to have a value between 0 and $10^{-4}(1/T_s)$, because a larger value of the frequency offset results in a huge increase of the PER. Recall that, as explained in Section \ref{sec:sec2}, $\Delta f$ varies between 0 and $10^{-2}(1/T_s)$ for the other curves. We can observe on the graph that for the same $E_s/N_0$ ratio, channel estimation using EM combined with autocorrelation gives lower PER than all the other estimation techniques. However, degradation compared to perfect channel state information (CSI) starts increasing at $E_s/N_0=1.6$ dB. Simulations done for this estimation method with the assumption of $\Delta f=0$ prove that the error on the estimated frequency offset $\widehat{\Delta f}$ is the main cause of performance degradation.

\subsection{Estimation using pilot symbol assisted modulation}
\label{ssec:subsec5}

\begin{figure}[!t]
\begin{minipage}[b]{1.0\linewidth}
  \centering
  \centerline{\includegraphics[scale=0.6]{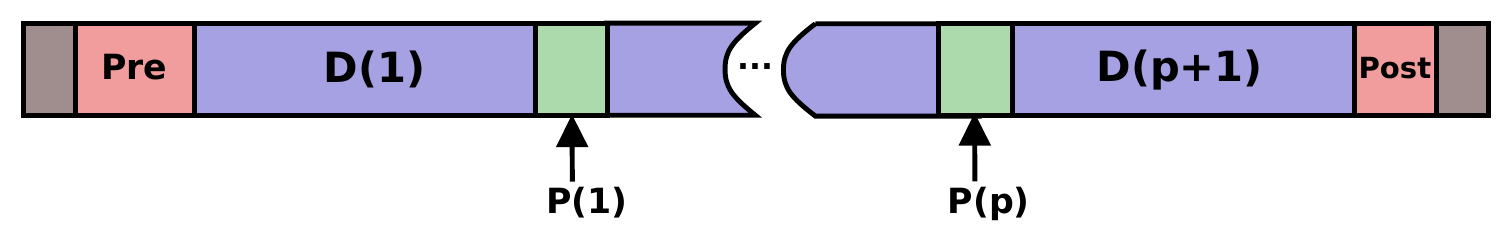}}
\end{minipage}
\caption{Packet structure with PSAM}
\label{fig:structure2}
\end{figure}

To refine the estimation of ${\Delta f}$, we use pilot symbol assisted modulation (PSAM) \cite{freqOff}. PSAM relies on the insertion of orthogonal data blocks called pilots inside the payload data sequence (see \figurename~\ref{fig:structure2}). Like the preamble and the postamble, pilots are Walsh-Hadamard words modulated with BPSK.

Vector $T$ in Equation \eqref{eq:EM_eq2} becomes: $T=\lbrace L_g+1,\ldots ,L_g+L_{pre}, L_g+L_{pre}+M+1,\ldots ,L_g+L_{pre}+M+L,\ldots ,L_g+L_{pre}+PM+(P-1)L+1,\ldots ,L_g+L_{pre}+P(M+L), L_g+L_{pre}+(P+1) M+PL+1,\ldots ,L_g+L_{pre}+(P+1) M+ P L+L_{post} \rbrace$.

\begin{figure}[!t]
\begin{minipage}[b]{1.0\linewidth}
  \centering
  \centerline{\includegraphics[scale=0.7]{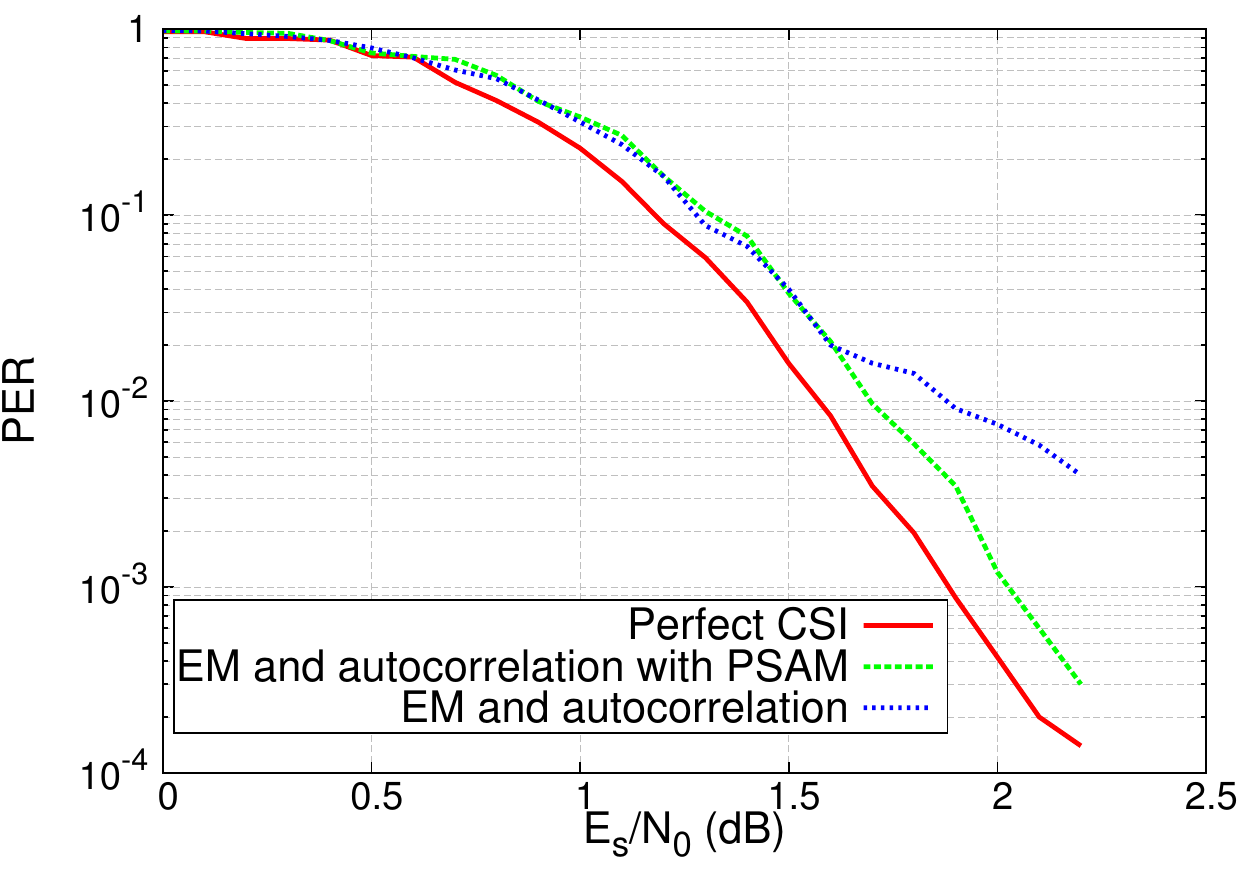}}
\end{minipage}
\caption{PER vs $E_s/N_0$ after interference cancellation and channel estimation using EM combined with autocorrelation and PSAM}
\label{fig:graph2}
\end{figure}

With PSAM, we are able to estimate the initial value of $\Delta f$ for the EM algorithm as follows
\begin{align}
\Delta f_k^{(0)}&=\frac{f_2(k)-f_1(k)}{2\pi(L+M)};\\
f_1(k)&=arg(r_{pre}\cdot s_{pre}(k)^T)\\
f_2(k)&=arg(r_{P1}\cdot s_{P1}(k)^T)
\end{align}
where $r_{P1}$ is the first pilot block of the received packet, $s_{pre}(k)$ and $s_{P1}(k)$ are the transpose vectors of the preamble of user $k$ and the first pilot block of user $k$, respectively. We uniformly distribute $9$ pilot blocks inside the packet, each of length equal to $12$ symbols. We reduce the preamble and postamble lengths to $40$ and $12$ symbols, respectively. Note that, PSAM induces a slight loss in the useful information rate compared to a packet structure with only a preamble and a postamble. The information rate loss is
\begin{equation}
Loss=\frac{S_{PSAM}-S}{S}
\end{equation}
where $S_{PSAM}$ is the packet size with pilot symbols and $S$ is the packet size without the pilots. With our parameters, the loss is equal to $5 \%$. The curve in Figure \ref{fig:graph2} corresponding to channel estimation using EM combined with autocorrelation and PSAM shows that the performance in presence of estimation errors is improved and the degradation of $E_s/N_0$ compared to perfect CSI is around $0.1$ dB.

\section{Discussion and conclusion}
\label{sec:sec5}

\begin{table}[!t]
\centering
  \begin{tabular}[c]{ |@{}>{\centering\arraybackslash}m{1.8cm}@{}|@{}>{\centering\arraybackslash}m{1.5cm}@{} |@{}>{\centering\arraybackslash}m{2cm}@{}|>{\centering\arraybackslash}m{3cm} |}
    \hline
    Autocorrelation \cite{casini} & EM \cite{EM2} & PSAM & Degradation in dB at $PER=10^{-3}$ \\ \hline
    \checkmark & & &$>0.5$ dB with $\Delta f\approx10^{-4}(1/T_s)$\\ \hline
    \checkmark & \checkmark & &$>0.5$ dB with $\Delta f\approx10^{-2}(1/T_s)$\\ \hline
    \checkmark & \checkmark &\checkmark  &$< 0.2$ dB with $\Delta f\approx10^{-2}(1/T_s)$\\ \hline
  \end{tabular}
\caption{Performance degradation in dB for different channel estimation techniques}
\label{table:summary}
\end{table}

In this paper, we have investigated existing channel estimation techniques used in recent random access methods. With the assumptions we have made, prior studies in \cite{casini} and \cite{EM2} still suffer from some limitations. We have proposed an improved channel estimation scheme that combines EM algorithm with autocorrelation estimation and pilot symbol assisted modulation. A brief comparison between the different estimation methods we have presented as well as the estimation scheme we have proposed is shown in Table \ref{table:summary}. Our work evaluated the effect of channel estimation errors in the case of just two superimposed packets.

Future work should consider the case of collision of several packets on the same time slot. Further, it should take into account imperfect symbol synchronization, phase noise impact and possible reduction of the computational complexity.

\bibliographystyle{IEEEbib}
\bibliography{Biblio2}

\end{document}